\documentclass[aps, reprint, onecolumn, 11pt, a4paper, longbibliography, nofootinbib, notitlepage]{revtex4-1}

\usepackage{amsthm, amsmath, amssymb, hyperref} \usepackage{color}
\usepackage[mathscr]{euscript}

\def\be{\begin{equation}}
\def\ee{\end{equation}}
\def\bea{\begin{eqnarray}}
\def\eea{\end{eqnarray}}
\def\bz{{\bar z}}

\def\p{\partial}
\def\bp{\bar\partial}

\def\bcal_k{\mathcal B_k}

\newcommand{\NPhi}{{N_{\phi}}}

\def\({\left(}
\def\){\right)}

\def\f{\mathsf{f}}

\def\g{{\rm g}}

\def\e{{\rm e}}
\def\b{{\rm b}}
\def\f{{\rm f}}
\def\L{{\rm L}}

\def\eq{Eq.\ \eqref}
\def\pf{{\rm Pf\,}}
\def\vz{{\mathbf{z}}}
\def\zs{{\left( \{ \vz_i \} \right) }}
\def\vw{{\mathbf{w}}}
\newcommand{\abs}[1]{\left| #1 \right|}
\newcommand{\norm}[1]{\left\lVert #1 \right\rVert}
\newcommand{\expt}[1]{\left\langle #1 \right\rangle}
\newcommand{\exptPf}[1]{\left\langle #1 \right\rangle_{\pf}}
\newcommand{\exptCFT}[1]{\left\langle #1 \right\rangle_{}}
\newcommand{\rmd}{\mathrm{d}}

\newtheorem*{maindefn*}{{\sc Definition}}

\newtheoremstyle{dotless}{}{}{\itshape}{}{\bfseries}{}{ }{}
\theoremstyle{dotless}

\newcommand{\vd}[1]{}
\renewcommand{\vd}[1]{{ \color{red} [VD: #1] }}

\begin{document}

\title{Geometric responses of the Pfaffian state}
\author{Vatsal Dwivedi}
\address{Institut f{\"u}r theoretische Physik, Universit{\"a}t zu K{\"o}ln, Zulpicher Stra{\ss}e 77a, 50937 K{\"o}ln, Germany} 
\email{vdwivedi@thp.uni-koeln.de}

\author{Semyon Klevtsov}
\address{Mathematisches Institut, Universit{\"a}t zu K{\"o}ln, Weyertal 86-90, 50931 K{\"o}ln, Germany}
\address{ITEP, Bol.~Cheremushkinskaya 25, 117259 Moscow, Russia}
\email{klevtsov@math.uni-koeln.de}

\begin{abstract} 
We define and study the Pfaffian state on Riemann surfaces with arbitrary metrics and  an inhomogeneous magnetic field and derive its universal transport coefficients. Following a path integral approach, we compute the generating functional which encodes the linear response of the system to a variation of the background metric and the magnetic field and use it to compute the leading and sub-leading corrections to the charge density in a large-$N$ expansion. We also present the first derivation of gravitational anomaly contribution at O$(k^6)$ to the static structure factor for the Pfaffian state in the long wavelength limit. 
\end{abstract}

\maketitle

\section{Introduction}

The study of universal features the fractional quantum Hall effect (FQHE) starting from explicit trial wave functions \cite{L83, H83,HR85,MR91} has led to a plethora of insights into the physics of interacting topological phases. Of particular interest are the transport coefficients, which are characteristic of the low energy excitations of the system. Interestingly, the low-energy spectrum of the collective excitations can be extracted by studying only the ground state of the quantum Hall droplet on the plane \cite{GMP86}. More generally, it has been shown that universal transport features such as the Hall conductivity and Hall viscosity can be extracted by considering the FQH ground state on two-dimensional compact manifolds \cite{WN90,ASZ94,ASZ95, L95,TV09,R09}.

A central object in the study of universal features of a FQH state is an effective action which encodes the response of the ground state to the background metric as well as the magnetic field. For many FQH states, an effective action has been proposed following general symmetry principles \cite{FS92,WZ92,Son13,AG14-2,BR15,GCYAF15,GS18}. However, in certain special cases, the action functional can be explicitly derived starting from the exact trial wave functions. Examples include the integer quantum Hall state for integer filling \cite{K14,KMMW17, AG14} and the much-studied Laughlin state for  $\nu = 1/q$, with $q$ being an odd integer \cite{CLW14,FK14,CLW15,LCW15,BR15-2,KW15,K16,K17}. 

A more intriguing trial wave function is the Pfaffian, proposed by Moore and Read \cite{MR91} for FQH where the filling fraction has an even denominator. In particular, the Pfaffian has been put forward as a strong candidate for the experimentally observed quantum Hall plateau at the filling fraction $\nu=5/2$. The recent measurement \cite{BHUFS18} of a fractional thermal Hall effect at this filling fraction has led to much debate on whether the Pfaffian state is actually realized \cite{WVH18, MOSMH18,S18,S18com,S18rep,ZF16}. Besides, the Pfaffian state has also been of particular theoretical interest owing to possible realization of nonabelian anyons, the braiding of which can be used to implement universal quantum computation \cite{NSSFD08}. The geometric and topological aspects of the Pfaffian state on Riemann surfaces have been an object of intense studies \cite{GWW92,GWW92-2,RR99,RG00,R09,RR11,H09,CS07,OKSNT07}. 

In this article, we compute the geometric response of the Pfaffian state on Riemann surfaces to arbitrary curved metrics and inhomogeneous magnetic fields. Our approach follows from the fact that the Pfaffian state can be written as a correlator in the Ising conformal field theory (CFT) where the electron operator consists of a noninteracting bosonic and a noninteracting Majorana fermionic operator. Generalizing the corresponding action functionals to curved spacetime backgrounds, we use a path integral formulation of the correlator to compute the generating functional in the large $N_\phi$ limit, where $\NPhi$ is the total flux of the magnetic field across the closed surface. The variational derivatives of the generating functional are then used to compute the observables. 

Our central results are the one- and two-point density correlators for the Pfaffian state with $B=N_\phi$ flux quanta  and filling fraction $\nu$. Using the one-point correlator, we compute the particle density in a large-$\NPhi$ expansion as
\be
\langle\rho(z, \bz)\rangle=\frac{\nu}{2\pi}\NPhi+\frac{\sqrt\nu Q}{8\pi}R+\frac1{8\pi\NPhi}\left[\frac{c}{12}+ \frac{\sqrt\nu Q}{4}\bigg(1-\frac1\nu\bigg)\right]\Delta_g R+\mathrm O(\NPhi^{-2}),
\ee
where $c = 3/2 = 1 + 1/2$ is the central charge of the Ising CFT, with the $1/2$ contribution from the Majorana fermion. Furthermore, $Q$ the background charge, $R$ the scalar curvature of the underlying manifold and $\Delta_g$ the corresponding Laplacian. Note that we have set $\hbar = e = 1$ and normalized the area of the underlying (compact) manifold to $2\pi$. The first two terms here encode the Hall conductivity and Hall viscosity, respectively. The central charge appearing in the third term is due to the \emph{gravitational (trace) anomaly} in the bulk. One way to see the connection between the $\Delta_gR$ term and the gravitational anomaly is via the variation of the Liouville action (\eq{GravAn}) as discussed in more detail in Refs \cite{K14,AG14-2,KMMW17}. This term has been derived for the integer \cite{K14} and fractional \cite{FK14,CLW14} quantum Hall states  starting from the trial wave function and independently from a $2+1$ dimensional effective field theory approach \cite{AG14,GCYAF15}. 

From the two-point correlator, we compute the static structure factor in the long wavelength limit as 
\be
S(k)=\frac{k^2}2+\frac{1-\nu}{8\nu}k^4+\frac{(1-2\nu)(2-\nu)}{64\nu^2}k^6+ \mathrm O(k^8).
\ee
More precisely this is an expansion in $(k\ell)^2$, 
written in the units where the magnetic length $\ell=\sqrt{\hbar/eB}$ is set to one.
Here, the gravitational anomaly first enters at $\mathrm O(k^6)$, a term which interestingly  vanishes for $\nu=1/2$ fermionic Pfaffian  state. We note that this gravitational anomaly contribution has been previously conjectured in Ref. \cite{CLW15} and in Ref. \cite{NCG17}, based on the extension of the Ward identity method of Ref. \cite{CLW15} to all chiral FQH states. This article thus presents the first explicit computation of the gravitational anomaly contribution to the static structure factor starting from the Pfaffian trial wave function.

The static structure factor for FQH states was first introduced by Girvin, Macdonald and Platzmann (GMP) \cite{GMP86}, where it was used to describe the so-called magneto-roton neutral excitation mode of density waves in the bulk of a quantum Hall droplet. These modes have been experimentally observed in inelastic light scattering \cite{PDPW93,KSSUK09} as well as numerically for various fractional quantum Hall effect \cite{MHKPS12,RNPR14,GNRS16,J17}. It has been suggested  \cite{GWW92} that the $\nu=5/2$ state exhibits, besides the GMP mode, another kind of gapped collective mode which has been termed the neutral fermion mode \cite{RG00,GS17,BFN11,MWC11,YHPH12,GMR19}.

The outline of the rest of this article is as follows: In Sec.~\ref{sec:bg}, we describe the CFT construction of the Pfaffian state on a plane and generalize it to arbitrary Riemann surfaces and inhomogeneous magnetic fields. In Sec.~\ref{sec:sph}, we specialize to the case of a sphere and compute the generating functional for the Pfaffian state for an arbitrary background metric. In Sec.~\ref{sec:corr}, we use the generating functional to compute various density correlators and hence the electron density and the static structure factor in a large-$N_\phi$ expansion. Finally, we discuss our results in Sec.~\ref{sec:conc}.

\section{Background and definitions} 
\label{sec:bg}

\subsection{FQHE-CFT connection}
We start off with a brief overview of the description of trial wave functions for FQHE on the plane $\mathbb{C}$ as conformal blocks of a CFT. The Laughlin state with filling fraction $\nu = 1/q, \, q \in \mathbb{Z_+}$ for a constant magnetic field $B$ is described by the wavefunction
\be
    \Psi_{\L} \zs = \frac{1}{\sqrt{Z_\L}} \prod_{i<j} (z_i - z_j)^q \; \exp\left\{- \frac{B}{4}\sum_{i=1}^N \abs{z_i}^2 \right\},
\ee
where $z_i \in \mathbb{C} $, $\mathbf{z} \equiv (z,\bz)$ and $Z_\L$ is a normalization constant. The Laughlin wave function can be written in terms of a correlation function in the free scalar boson CFT, \cite{DMS} consisting of a single scalar field $\varphi(z, \bar{z})$ described by the action \footnote{
    Note that this action conformally invariant only for $B=0$, since the term linear in $B$ breaks the conformal invariance by introducing a length scale, \emph{viz}, the magnetic length, in the system. 
}
\begin{equation}
    S_\b[\varphi] 
    = \frac{1}{2\pi} \int_{\mathbb{C}}  \rmd^2z \left( \frac12 \p_z\varphi \, \p_{\bz}\varphi + i\sqrt{\nu} \, B \varphi  \right).
    \label{boson_action1}
\end{equation}
The second term in this action, linear in $\varphi$, is required in order to satisfy the charge neutrality condition. The correlation function of a string of operators $\mathcal{O}(\vz_i)$ in the CFT is defined as 
\be 
    \exptCFT{ \mathcal{O}(\vz_1)\cdots \mathcal{O}(\vz_N) }  = \int \mathcal{O}(\vz_1)\cdots \mathcal{O}(\vz_N)\, \e^{-S_\b[\varphi]} \; \mathcal D \varphi.
\ee  
The modulus squared of the Laughlin wave function can then be written as an expectation value of a string of electron operators $\mathcal{O}_e(\vz) = \e^{i\varphi(\vz)/\sqrt{\nu}}$. Explicitly, up to a normalization constant $C$,
\be
    \abs{ \Psi_{\L} \zs}^2 =C \exptCFT{ \mathcal{O}_e(\vz_1)\cdots \mathcal{O}_e(\vz_N) }.
    \label{L_cft}
\ee 
This can be derived using the two point correlation function
\be
    \exptCFT{ \varphi(\vz) \varphi(\vz') } = -\log (z-z') - \log (\bz-\bz'),
\ee 
which can be computed explicitly from the action. 

The Pfaffian wave function on the plane, first introduced by Moore-Read in Ref.~\cite{MR91}, can be written for an even number of particles as 
\be
    \Psi_\pf \zs = \frac{1}{\sqrt{Z_{\rm Pf}}} \, \pf \left( \frac{1}{z_n-z_m} \right) \, \prod_{i<j} (z_i - z_j)^q \; \exp\left\{- \frac{B}{4}\sum_{i-1}^N \abs{z_i}^2 \right\},
\ee
where $Z_{\rm Pf}$ is a  normalization constant and Pf$(M) = \sqrt{\det M}$ denotes the Pfaffian of the $N$-dimensional antisymmetric matrix $M$ with $M_{nm} = 1/(z_n-z_m)$, which is defined only if $N$ is even and is always a polynomial in the matrix entries. The Pfaffian of an antisymmetric matrix $M$ can alternatively be defined directly in terms of the matrix elements as 
\begin{equation}
    \text{Pf}(M) \equiv \frac{1}{ 2^{N/2} \, (N/2)!} \sum_{\sigma \in S_{N}} \left[ \text{sgn}(\sigma) \prod_{i=1}^{N/2} M_{\sigma(2i-1),\sigma(2i)} \right] , 
\end{equation}
where $S_{N}$ is the group of permutations of $N$ symbols. The Pfaffian wave function can be expressed as a correlator in the Ising CFT \cite{MR91}, which, besides the
scalar boson, has Majorana fermionic fields $\psi, \bar\psi$ with the action
\be\label{Sf}
    S_\f[\psi,\bar\psi]
    = \frac{1}{2\pi}\int_{\mathbb{C}}\left( \psi\p_{\bz} \psi+\bar\psi \p_z \bar\psi \right)  \rmd^2z .
\ee
The modulus of the Pfaffian wave function can then be written as an expectation value:
\be
    \abs{ \Psi_\pf \zs }^2 = 
  C  \int \mathcal O_e(\vz_1)\cdots \mathcal O_e(\vz_N) \, e^{-S_\b[\varphi]-S_\f[\psi,\bar\psi]} \, \mathcal D\varphi\mathcal D\psi\mathcal D\bar\psi,
\ee
where the electron operators are now given by
\be 
\mathcal{O}_e(\mathbf{z}) = \norm{\psi}^2 \e^{i \varphi/\sqrt{\nu}} \,  \Big|_{\vz}, \qquad \norm{\psi(\mathbf{z})}^2 = \bar\psi(\mathbf{z}) \psi(\mathbf{z}).
\ee 
To explicitly derive this, we also need the fermionic correlation functions
\be
    \exptCFT{\psi(\vz)\psi(\vz')} = \frac{1}{z-z'}, \qquad 
    \exptCFT{\bar\psi(\vz)\bar\psi(\vz')} = \frac{1}{\bz-\bz'}, \qquad 
    \exptCFT{\psi(\vz)\bar\psi(\vz')} = 0,
\ee
which can again be computed explicitly from the fermionic action.

\subsection{Action functionals on Riemann surfaces}
We next consider the CFTs discussed above on an arbitrary Riemann surface $\Sigma$ with a conformal metric $g$, in order to generalize the Pfaffian state to arbitrary Riemann surfaces. Using local complex coordinates $(z, \bz)$ on $\Sigma$, the length element is given by $ \rmd s^2=2g_{z\bz}dzd\bz$ with $g_{z\bz}(\vz)$ a real-valued positive function on $\Sigma$. Then the scalar curvature can be written using the scalar Laplacian as 
\be 
    R = - \Delta_g \log g_{z\bz}, \qquad \Delta_g = \frac4{\sqrt g}\p_z\p_\bz,
\ee 
where $\sqrt{g} = 2g_{zz}$. The bosonic action on $\Sigma$ is a generalization of Eq.~\ref{boson_action1} to curved backgrounds:
\begin{equation}
\label{action}
    S_\b[\varphi] = \frac1{2\pi}\int_\Sigma\Big(\frac12g^{z\bz}\p_z\varphi\p_{\bz}\varphi + i\sqrt\nu B \varphi + \frac{i}{4} Q R \, \varphi  \Big)\sqrt g \,  \rmd^2z,
\end{equation}
where the magnetic field $B(\mathbf{z})$ is allowed to be inhomogeneous. We are also free to couple $\varphi$ to the scalar curvature $R$ with the coupling constant $Q$, which is usually termed the \emph{background charge}. We compactify the scalar field on a circle of radius $R_c$ by the identification $\varphi\sim\varphi+2\pi R_c$, which can lead to winding field configurations if $H^1(\Sigma, \mathbb{Z}) \neq 0$. We have the following global constraints
\begin{equation}\label{quant}
    \frac{1}{2\pi} \int_\Sigma \sqrt{g} \,  \rmd^2z = 1, \qquad 
    \frac{1}{2\pi} \int_\Sigma B \, \sqrt{g} \,  \rmd^2z = \NPhi, \qquad 
    \frac{1}{4\pi} \int_\Sigma R \, \sqrt{g} \,  \rmd^2z = \chi.
\end{equation}
The first equation is the normalization of the area of $\Sigma$, the second is a statement of constant flux with $\NPhi \in \mathbb{Z}_+$ the number of flux quanta and the third is the Gauss-Bonnet theorem with the Euler character $\chi = 2-2\g$, where $\g$ is the genus of $\Sigma$. 

The fermionic action on $\Sigma$ is a generalization of \eq{Sf}:
\be
    S_\f [\psi,\bar\psi] = \frac1{2\pi}\int_\Sigma (\psi\p_\bz\psi+\bar\psi\p_z\bar\psi)  \rmd^2z = \frac1{2\pi}\int_\Sigma (\psi\bar\p\psi+\bar\psi\p\bar\psi).
\ee
The latter definition is coordinate-independent, but uses the machinery of derivatives on fiber bundles.  Recall that the fermionic field $\psi$, resp.\ $\bar\psi$, transforms as a section of a \emph{spin bundle} of choice $\mathcal S$, resp.\ its complex conjugate $\bar{\mathcal S}$, over $\Sigma$. One can define  Dolbeault operators $\p, \bar\p$ which act as derivatives on these fields. More formally, $\bp$ acts as a derivative on smooth sections of $\mathcal{S}$, the space of which is denoted by $C^\infty(\Sigma,S)$, so that we can write \footnote{
    Often the notation $\bp_{L}$ is used for the del bar operator for a bundle $L$, see e.g. D'Hoker-Phong VI.C, \cite[Eq.\ 24]{KMMW17}, while here we simply use short hand notation $\bp$.
}
\be\label{dbar}
    \bp: \mathcal C^\infty(\Sigma, \mathcal S)\to \Omega^{(0,1)}(\Sigma, \mathcal S),
\ee
where $\Omega^{(0,1)}(\Sigma, \mathcal S)$ is the space of $(0,1)$ forms\footnote{ 
    A $(m,n)$ differential form on a complex manifold with complex dimensions $d \geq m,n$ corresponds to $\rmd z_1 \wedge \dots \wedge \rmd z_m \wedge \rmd \bz_1 \wedge \dots \wedge \rmd\bz_n$. 
}. 
In local coordinates $(z,\bz)$, given a fermionic field $\psi(\vz)$, this action is simply $\bp \psi = \p_\bz\psi\, \rmd\bz$. Thus, $\psi\bp\psi\in S\otimes \Omega^{(0,1)}(\Sigma, \mathcal S)\sim\Omega^{(1,1)}(\Sigma)$ is a $(1,1)$ form that can be integrated over $\Sigma$. The identification is due to the fact that the square of a spin bundle $\mathcal S^2=K$ equals the canonical line bundle $K$ on $\Sigma$, which is a bundle of $(1,0)$-forms. Hence one can think of $\psi$ as a half-form on $\Sigma$ \cite{DP88}.

\subsection{The Pfaffian state on Riemann surfaces}
For the Pfaffian state on $\Sigma$, the electron operator takes the form 
\be
\label{elop}
    \mathcal O_e(\mathbf{z})=\norm{\psi}^2 \cdot e^{i\varphi/\sqrt\nu} \Big|_\mathbf{z}; \qquad
    \norm{\psi(\vz)}^2 = g_{z\bz}^{-1/2}(\vz) \bar\psi(\vz) \psi(\vz).
\ee
We need the additional factor of $g_{z\bz}^{-1/2}$ -- the Hermitian metric on the spin bundle $\mathcal S$ -- in the definition of $\norm{\psi}$ so that the full operator transforms like a scalar field on $\Sigma$. The electron operator is a primary field with scaling dimension\footnote{ 
    This is often termed the conformal, or gravitational, spin \cite{CLW14,FK14}, see also \cite{R09}.
}
\be\label{confdim}
    s = \frac12 + \frac1{2\nu} \left( 1 - Q \sqrt{\nu} \right),
\ee
where the one-half is the fermionic contribution and the rest is due to the boson. As in the planar case, we consider the correlator 
\be
\label{exp}
    \mathcal V(g,B,\{\vz_i\})
    = \int \mathcal O_e(\vz_1)\cdots \mathcal O_e(\vz_N)\,e^{-S_\b(\varphi)-S_\f(\psi,\bar\psi)}\mathcal D_g\varphi\mathcal D\psi\mathcal D\bar\psi.
\ee
Demanding a nontrivial value for $\mathcal V(g,B,\{\vz_i\})$ leads to constraints on the parameters
$N,\NPhi,s$ and $Q$. To derive these, recall that the scalar field admits a mode expansion $\varphi(\mathbf{z}) = \varphi_0 + \tilde\varphi(\mathbf{z})$, where $\tilde\varphi(\mathbf{z})$ has mean zero on $\Sigma$ \cite{DMS}. For the zero mode, the bosonic action becomes 
\be 
    S_\b[\varphi_0] = \frac{i\varphi_0}{2\pi} \int_\Sigma \left( \frac{Q}4 R + \sqrt\nu B \right) \sqrt g \,  \rmd^2z 
    = i \varphi_0 \, \left[ \frac{Q \,\chi}{2}  +  \NPhi\sqrt{\nu}  \right].
\ee 
We can perform the integral over $\varphi_0$ in \eq{exp} to get 
\be\label{zeromode}
    \int  \rmd\varphi_0 \; e^{-S[\varphi_0]} \, \prod_{i=1}^N \e^{i \varphi_0/\sqrt{\nu}} \;
    = \int  \rmd\varphi_0 \; \exp\left\{i\left( \frac{N}{\sqrt{\nu}} - \frac{Q \,\chi}{2} - N_\phi \sqrt{\nu} \right) \varphi_0 \right\}.
\ee
For this integral to be nonzero, we demand that 
\be \label{N}
    N = \nu \left[ \NPhi +  \frac{Q \,\chi}{2\sqrt\nu } \right] =\nu \left[ \NPhi + \left( \frac{1}{\nu} + 1-2s \right) \frac{\chi}{2} \right].
\ee 
where we have used \eq{confdim} to write 
\be
Q=\frac1{\sqrt\nu}+(1-2s)\sqrt\nu.
\ee
Physically, \eq{N} represents a charge conservation constraint, while mathematically, it is analogous to the Riemann--Roch theorem\footnote{
    More precisely, when $\nu=1$, this relation is equivalent to the Riemann--Roch theorem for the line bundle $L\otimes K^{s-\frac12}$, where $L$ is a degree-$\NPhi$ magnetic line bundle.
}. 
% Since the filling factor on a plane is defined by the relation $N_\phi = N/\nu$, it is customary to call the deviation from this expression as the the shift parameter \cite{WZ92} $\mathscr S$, defined by $\NPhi=N/\nu-\mathscr S$.  
Finally, since $\varphi$ is compactified, the partition function and hence the integral in \eq{zeromode} should be invariant under a translation $\varphi\sim\varphi+2\pi R_c$, which requires that
\be 
   \left[ \frac{N}{\sqrt{\nu}} - \frac{Q \,\chi}{2} - N_\phi \sqrt{\nu} \right] R_c  
   = \left[ N - Q \sqrt{\nu} (1-\g) - \nu N_\phi \right] \frac{R_c}{\sqrt{\nu}} \in\mathbb Z,
\ee
We shall assume that $\nu\NPhi$ and $Q\sqrt \nu$ are integers and $s\in\mathbb{Z}/2$, so that we may choose $R_c = 1/\sqrt{\nu}$. 

We now explain the relation between the wave function and the correlator \eq{exp}. 
Unlike the case of the plane, $\mathcal V(g,B,\{z_i\})$ is not quite $\abs{\Psi_\pf}^2$ if $H^1(\Sigma, \mathbb{Z}) \neq 0$. This is because in presence of non-contractible loops on $\Sigma$, the definition of the right hand side involves the sum over all choices of the spin structure, i.e, the choice of periodic/anti-periodic boundary condition for fermions for each non-contractible loop. Thus, we get a set of degenerate Pfaffian states, which satisfy
\be
   \frac{1}{n_\g} \sum_{n=1}^{n_{\g}} \abs{\Psi_{\pf,n} \zs}^2 = \frac1{Z[g,B]} \; \mathcal V(g,B,\{\vz_i\}), 
\ee
where the degeneracy $n_{\g}$ corresponds to the number of odd/even spin structures \cite{RG00, OKSNT07}. The normalization constant is defined as 
\be\label{normlz}
  Z_\pf[g, B] = \int_{\Sigma^N} \mathcal{V}(g,B, \{\vz_i\}) \, \prod_{i=1}^N \sqrt{g} \,  \rmd^2z_i.
\ee
The normalization constant might seem like an unimportant overall constant, but for fractional quantum Hall states on curved backgrounds, its variation with respect to $B$ and $g$ can be used to compute various transport coefficients \cite{FK14,CLW15}, as we show in Sec.~\ref{sec:corr}.

\section{The Pfaffian state on the sphere}
\label{sec:sph}

\subsection{Correlators}
In this section, we explicitly compute the Pfaffian state on a Riemann surface of genus zero, i.e, a sphere. This state is unique so that $\abs{\Psi_\pf\zs}^2 \propto \mathcal V(g,B,\{\vz_i\})$ and the correlator splits into a product of bosonic and fermionic terms as
\begin{align}
\mathcal V(g,B,\{\vz_i\}) &= \int \norm{\psi(\vz_1)}^2\cdots \norm{\psi(\vz_i)}^2\,\e^{-S_\f[\psi,\bar\psi]}\mathcal D\psi\mathcal D\bar\psi
\cdot \int \e^{\frac{i}{\sqrt{\nu}}\varphi(\vz_1)}\cdots \e^{\frac{i}{\sqrt{\nu}}\varphi(\vz_N)}\,\e^{-S_\b[\varphi]}\mathcal D_g\varphi, \nonumber \\ 
&= \expt{\norm{\psi(\vz_1)}^2\cdots \norm{\psi(\vz_N)}^2}_\f \cdot \expt{\e^{\frac{i}{\sqrt{\nu}}\varphi(\vz_1)}\cdots \e^{\frac{i}{\sqrt{\nu}}\varphi(\vz_N)}}_\b.
\end{align}
We begin by noting that there are no zero mode integrations for either bosonic or fermionic path integrals. The former follows from the fact that there are no noncontractible loops on the sphere, while the latter from the fact that there are no holomorphic sections of $\mathcal S$ on the sphere \cite[Eq.\ 2.53]{DP88}, so that the operator $\bp$ defined in \eq{dbar} has $\ker\bar\p=0$. In the following, we compute the two correlators separately.

\subsubsection{The bosonic correlator}
This is simply the Laughlin state with filling fraction $\nu$ \cite{FK14,K16}, whose computation follows from standard quantum field theory techniques~\cite{DMS}. Explicitly, we can write the bosonic correlator as 
\begin{align}                
    & \expt{\e^{\frac{i}{\sqrt{\nu}}\varphi(\vz_1)}\cdots \e^{\frac{i}{\sqrt{\nu}}\varphi(\vz_N)}}_\b \nonumber \\ 
    & \qquad\qquad =  \int \exp\left\{  - \frac{1}{2\pi} \int_\Sigma \sqrt{g} \,  \rmd^2z \left( \frac12 g^{z\bz} \p_z\varphi\p_{\bz}\varphi + i\sqrt{\nu} \, B \varphi + \frac{iQ}{4} \, R \, \varphi \right) + \frac{i}{\sqrt{\nu}} \sum_{i=1}^N \varphi(\vz_i)  \right\}  \mathcal D_g\varphi, \nonumber \\ 
    & \qquad\qquad =  \int \exp\left\{  -\frac{1}{2\pi} \int_\Sigma \sqrt{g} \,  \rmd^2z \left( \frac14 \varphi (-\Delta_g) \varphi - i J \varphi \right)  \right\}  \mathcal D_g\varphi,
\end{align}
where we have integrated the first term by parts and grouped the remaining terms linear in $\phi$ by defining the source 
\be
    J(\mathbf{z}) = \frac{2\pi}{\sqrt\nu} \sum_{i=1}^N \delta_g (\vz, \vz_i) - \sqrt\nu B(\mathbf{z}) - \frac{Q}{4} R(\mathbf{z}) .
\ee
Here, $\delta_g(\vz,\vz')$ is the Dirac delta `function' on $\Sigma$, which satisfies $\int_\Sigma \delta_g(\vz,\vz') f(\vz) \sqrt{g} \, \rmd^2z = f(\vz')$ for any smooth function $f$. The path integral can now be evaluated by a linear shift 
\be 
    \varphi \to \varphi + \frac{i}{\pi} \int_\Sigma G^g(\vz,\vz') J(\vz') \sqrt{g} \,  \rmd^2z', 
\ee
where $G^g$, the Green's function for the scalar Laplacian, is defined by\footnote{ 
    We need the $-1$ on the right hand side of this definition since the Laplacian on a sphere has a zero mode, \emph{viz}, the constant function. 
}
\be\label{Gdef}
    -\Delta_g G^g(\vz,\vz') = 2\pi \delta_g(\vz,\vz') - 1, \qquad 
    \int_\Sigma  G^g(\vz,\vz') \sqrt{g} \,  \rmd^2z'= 0. 
\ee
Thus, the bosonic correlator becomes
\be
 \exp\left\{ -\frac{1}{4\pi^2} \int_\Sigma \sqrt{g} \,  \rmd^2z \sqrt{g} \,  \rmd^2z' \, J(\vz) G^g(\vz,\vz') J(\vz')  \right\}  \int \exp\left\{  -\frac{1}{8\pi} \int_\Sigma \sqrt{g} \,  \rmd^2z \, \varphi (-\Delta_g) \varphi \right\}  \mathcal D_g\varphi.
\ee
The remaining path integral is simply a Gaussian integral, which evaluates to the regularized determinant of the Laplacian. Substituting the explicit form of $J(\vz)$, we finally get (also see \cite[Eq.\ 4.26]{K16})
\begin{align}\label{Vbos}\nonumber
& \left[\frac{\det'\Delta_g}{2\pi}\right]^{-1/2}\cdot \exp\left\{-\frac1{4\pi^2}\int_{\Sigma\times\Sigma}\left(\frac Q4R+\sqrt\nu B\right)\bigg|_\vz G^{g
}(\vz,\vz')\left(\frac Q4R+\sqrt\nu B\right)\bigg|_{\vz'}\sqrt{g} \, \rmd^2z\,\sqrt{g} \, \rmd^2z'\right\} \\
& \cdot\exp \left\{ \frac1{\pi\sqrt\nu} \sum_{i=1}^{N}\int_\Sigma G^{g}(\vz_i,\vz)\left(\frac Q4R+\sqrt\nu B\right)\bigg|_\vz\sqrt{g} \rmd^2z-\frac1\nu\sum_{i\neq j}^{N}G^{g}(\vz_i,\vz_j)-\frac1\nu\sum_{i=1}^{N}G^{g}_{\rm reg}(\vz_i)\right\},
\end{align}
where $\det'$ indicates the product of eigenvalues excluding the zero mode, and $G^g_{\rm reg}$ is the regularized Green function at coincident points, defined as 
\be
\label{reggreen}
    G^{g}_{\rm reg}(\vz)=\lim_{\vz\to \vz'} \left( G^{g}(\vz,\vz') + \log d_{g}(\vz,\vz') \right),
\ee
where $d_{g}(\vz,\vz')$ is the geodesic distance between the points in the metric $g$.

\subsubsection{The fermionic correlator} 
This is the new contribution to the Pfaffian state in comparison to the Laughlin state. The computation essentially follows from the fact that for free fermions, we can use Wick's theorem to write the full correlator as a sum over pairwise correlators (propagators) $\langle\psi(\vz_i)\psi(\vz_j)\rangle$ and $\langle\bar\psi(\vz_i)\bar\psi(\vz_j)\rangle$. Geometrically, these are sections of $\mathcal S$, i.e., half forms in both $\vz_i$, $\vz_j$ and antisymmetric on $\Sigma \times \Sigma$, which can be written with some abuse of notation as  
\begin{equation}\label{psipsi}
\langle\psi(\vz_i)\psi(\vz_j)\rangle=\mathcal P(\vz_i,\vz_j)\sqrt{\rmd z_i}\sqrt{\rmd z_j},
\end{equation}
The defining equation is analogous to the Green's function equation for the bosonic case:
\be
\langle\bp\psi(\vz_i)\psi(\vz_j)\rangle=\delta_g (\vz_i,\vz_j) \rmd z_i\wedge \rmd \bz_j
\ee
The left hand side is a $(\frac12,1)$-form in $z_i$ and a $(\frac12,0)$-form in $z_j$, thereby combining into a $(1,1)$-form. We can thus write 
\begin{equation}\label{prop}
g^{z\bz}\p_{\bz_i}\mathcal P(\vz_i,\vz_j)=2\pi\delta_g(\vz_i,\vz_j).
\end{equation}
The solution on the sphere is given by
\be
\mathcal P(\vz_i,\vz_j)=\frac1{z_i-z_j}
\ee
and $\langle\psi(\vz_i)\psi(\vz_j)\rangle$ is the inverse of the Prime-form on the sphere \cite[p.3.207]{Mumford}.
Since \eq{prop} is invariant under diffeomorphisms, 
% the transformation of metric within the conformal class $g_{z\bz}\to e^{\sigma(\mathbf{z})}g_{z\bz}$, 
the function $\mathcal P(\vz_i,\vz_j)$ is metric independent, unlike the Green's function for the bosonic case. The fermionic correlation function reads
\begin{align}\label{Vferm}
    \expt{\norm{\psi(\vz_1)}^2\cdots \norm{\psi(\vz_N)}^2}_\f
    % &= \int \norm{\psi(\vz_1)}^2\cdots \norm{\psi(\vz_N)}^2\,\e^{-S_\f[\psi,\bar\psi]}\mathcal D\psi D\bar\psi \nonumber \\
    &= \prod_{i=1}^N g_{z\bz}(\vz_i)^{-1/2} \expt{\psi(\vz_1) \cdots \psi(\vz_N)} \cdot 
    \expt{\bar\psi(\vz_1) \cdots \bar\psi(\vz_N)}  \nonumber \\ 
    &= {\pf}\,\bp\cdot{\pf}\,\p\cdot\prod_{i=1}^Ng_{z\bz}(\vz_i)^{-1/2}\cdot\big|\pf \mathcal P(\vz_n,\vz_m)\big|^2.
\end{align}
% 
% \begin{multline}\label{Vferm}
% \int ||\psi(\vz_1)||^2\cdots ||\psi(\vz_N)||^2\,e^{-S_\f}\mathcal D\psi D\bar\psi\\=\prod_{i=1}^N g_{z\bz}(\vz_i)^{-1/2} \cdot\langle\psi|_{\vz_1}\cdots \psi|_{\vz_N}\rangle\langle\bar\psi|_{\vz_1}\cdots \bar\psi|_{\vz_N}\rangle\\
% ={\pf}\,\bp\cdot{\pf}\,\p\cdot\prod_{i=1}^Ng_{z\bz}(\vz_i)^{-1/2}\cdot\big|\pf \mathcal P(\vz_n,\vz_m)\big|^2
% \end{multline}
% 
The product of two Pfaffians above can be rewritten as \cite[Eq.\ 10.32]{DMS}
\be\label{pfdet}
{\pf}\,\bp\cdot{\pf}\,\p=\left[\det\bp^\dagger\bp\right]^{1/4}\left[\det\p^\dagger\p\right]^{1/4},
\ee
where the adjoint operators are defined with respect the inner product
\be\label{inner}
\langle\psi_1,\psi_2\rangle=\int_\Sigma g_{z\bz}^{-1/2} (\vz)  \psi_1^*(\vz) \psi_2(\vz) \sqrt g \, \rmd^2z.
\ee
The adjoint of $\bp$ has no zero modes on the sphere, i.e., $\ker\bp^\dagger=0$ \cite[Eq.\ 2.52]{DP88}. 
% where the notations are $\bp=\nabla^z_{1/2}$ and $\bp^\dagger=\nabla^{-1/2}_z$ and $\det\bp^\dagger\bp=\Delta^{(-)}_{1/2}$.

Putting together \eq{Vbos} and \eq{Vferm}, we get $\abs{\Psi_\pf}^2$ on the 2-sphere with an arbitrary metric $g$ as 
\begin{align}\label{Vgen}\nonumber
& \mathcal V(g,B,\{\vz_i\})=\left[\frac{\det'\Delta_g}{2\pi}\right]^{-1/2}[\det\bp^\dagger\bp\cdot\det\p^\dagger\p]^{1/4}\cdot\prod_{i=1}^Ng_{z\bz}(\vz_i)^{-1/2}\cdot\abs{\pf \left( \frac{1}{z_n-z_m} \right)}^2\\
& \quad \times  \exp\left\{-\frac1{4\pi^2}\int_{\Sigma\times\Sigma}\left(\frac Q4R+\sqrt\nu B\right)\bigg|_\vz G^{g
}(\vz,\vz')\left(\frac Q4R+\sqrt\nu B\right)\bigg|_{\vz'}\sqrt{g} \, \rmd^2z\,\sqrt{g} \, \rmd^2z'\right\} \\
& \quad\times\exp \left\{ \frac1{\pi\sqrt\nu} \sum_{i=1}^{N}\int_\Sigma G^{g}(\vz_i,\vz)\left(\frac Q4R+\sqrt\nu B\right)\bigg|_\vz\sqrt{g} \,  \rmd^2z - \frac1\nu\sum_{i\neq j}^{N}G^{g}(\vz_i,\vz_j)-\frac1\nu\sum_{i=1}^{N}G^{g}_{\rm reg}(\vz_i)\right\}. \nonumber 
\end{align}

\subsection{The round sphere}
To illustrate the above description of the Pfaffian state, we now specialize it for the case of the \emph{round metric} on the 2-sphere, which is explicitly given by
\begin{equation}\label{roundsp}
g_{0,z\bz}=\frac{1}{(1+|z|^2)^2}.
\end{equation}
This metric has a constant scalar curvature $R = 4$. The Green's function for the Laplacian is given by
\begin{align}\nonumber
G^{g_0}(\vz,\vz') =-\log\frac{|z-z'|}{\sqrt{(1+|z|^2)(1+|z'|^2)}}-\frac12, \qquad G_{\rm reg}^{g_0}(\vz) =-\frac12.
\end{align}
For a constant magnetic field $B_0=\NPhi$ we take the magnetic potential $h_0$ as,
\be
    B_0=-g_0^{z\bz}\p_z\p_{\bz}\log h_0^{\NPhi}(\mathbf{z}),\qquad h_0(\vz)=\frac1{1+|z|^2}.
\ee
To evaluate \eq{Vgen}, we begin by noting that since $B$ and $R$ are constants, the integrals over $G(\vz, \vz')$ vanish using the definition of Green's function from \eq{Gdef}. Furthermore, 
\be\nonumber 
\sum_{i\neq j}^{N}G^{g}(\vz_i,\vz_j) + \sum_{i=1}^{N}G^{g}_{\rm reg}(\vz_i) = -\log \left[ \prod_{i\neq j} \abs{z_i-z_j} \cdot \prod_{i=1}^N \frac{1}{(1 + \abs{z_i}^2)^{N-1}} \right] - \frac{N^2}{2}.
\ee
From \eq{N}, we also get $N = \nu (N_\phi -2s + 1) + 1$. Thus, 
\begin{align}
\mathcal V\bigl(g_0,B_0,\{\vz_i\}\bigr)=C_0\,\abs{\pf \left( \frac{1}{z_n-z_m} \right)}^2\ \cdot\prod_{i<j}^N|z_i-z_j|^{2/\nu}\cdot\prod_{i=1}^N\frac1{(1+|z_i|^2)^{\NPhi-2s}},
\end{align}
where $C_0$ is a numerical constant independent of $\vz_i$.  In the literature (for instance, \cite[Eq.\ 2.1]{GWW92-2}), the standard relation  for $\nu=1/2$ state on the sphere is $\NPhi=2N-3$, which corresponds to setting $s=0$ above. 

The wave function can alternatively be written as
\begin{align}\label{pf1}
\mathcal V\bigl(g_0,B_0,\{\vz_i\}\bigr)=C_0\,|F_{\pf}(z_1,...,z_N)|^2\cdot\prod_{i=1}^Nh_0^\NPhi(\vz_i)\, g_{0,z\bz}^{-s}(\vz_i),
\end{align}
where the
holomorphic part is 
\be\label{pfhol}
F_{\pf}(z_1,...,z_N)=\pf\left(\frac1{z_i-z_j}\right)\cdot\prod_{i<j}^N(z_i-z_j)^{1/\nu}
\ee 
and the non-holomorphic part indicates that $\mathcal V$ corresponds to magnetic flux $\NPhi$ and conformal spin $s$. The form of \eq{pf1} also holds for other choices of the reference metric. The expression in \eq{pf1} as it stands is unnormalized, so that the actual wave function is
\be
|\Psi_{\pf}(\vz_1,...,\vz_N)|^2 = \frac{C_0}{Z[g_0,B_0]}\,|F_{\pf}(z_1,...,z_N)|^2\cdot\prod_{i=1}^Nh_0^\NPhi(\mathbf{z}_i)\, g_{0,z\bz}^{-s}(\mathbf{z}_i)
\ee
with the normalization constant explicitly given by
\begin{align}
Z[g_0.B_0]& = C_0 \int_{\Sigma^N}|F_{\pf}(z_1,...,z_N)|^2\cdot\prod_{i=1}^Nh_0^\NPhi(\mathbf{z}_i)\, g_{0z\bz}^{-s}(\mathbf{z}_i)\prod_{i=1}^N\sqrt{g_0}  \, \rmd^2z_i \nonumber \\
&=\int_{\Sigma^N}\mathcal V\bigl(g_0,B_0,\{\vz_i\}\bigr)\prod_{i=1}^N\sqrt{g_0} \,  \rmd^2z_i.
\end{align}
In the following, we shall use this expression without referring to an explicit form of $g_0$.

\section{Computing the observables}
\label{sec:corr}

\subsection{The generating functional} 
The variation of the normalization constant with respect to the background metric can be used to compute the density correlation functions for the Pfaffian state. We consider an arbitrary metric $g_{z\bz}(\vz) =e^{\sigma(\mathbf{z})} g_{0z\bz}(\vz)$ in a conformal class of a reference metric $g_0$. An arbitrary positive magnetic field can be written as
\be\label{B}
B(\mathbf{z})=-g^{z\bz}\p_z\p_{\bz}\log h^\NPhi(\mathbf{z}),\quad h(\mathbf{z})=h_0(\mathbf{z}) e^{-\lambda(\mathbf{z})},
\ee
which is parametrized by the relative magnetic potential $\lambda(\mathbf{z})$ such that $B>0$ everywhere on $\Sigma$. The normalized wave function under the modified metric reads
\be\label{Psi2}
|\Psi_{\rm Pf}(\vz_1,...,\vz_N)|^2 = \frac{C_0}{Z[g_0,B_0,\sigma,\lambda]}\,|F_{\pf}(z_1,...,z_N)|^2\cdot\prod_{i=1}^Nh^\NPhi(\mathbf{z}_i)\, g_{z\bz}^{-s}(\mathbf{z}_i),
\ee
with the $\sigma$- and $\lambda$-dependent normalization constant 
\be\label{Z0}
Z[g_0,B_0,\sigma,\lambda] = C_0\int_{\Sigma^N}|F_{\pf}(z_1,...,z_N)|^2\cdot\prod_{i=1}^Nh^\NPhi(\mathbf{z}_i)\, g_{z\bz}^{-s}(\mathbf{z}_i)\prod_{i=1}^N\sqrt{g} \, \rmd^2z_i.
\ee
We now define the generating functional \cite{K14,FK14} as 
\begin{equation}\label{ZZ}
\log\frac{Z[g_0,B_0,\sigma,\lambda]}{Z[g_0,B_0]}=-S_{\rm eff}[g_0,B_0,\sigma,\lambda]+\mathcal F[g,B]-\mathcal F[g_0,B_0],
\end{equation}
where we have introduced the notation
\begin{align}\label{F}
\mathcal F[g,B]&=\log  \int_{\Sigma^N}\mathcal V\big(g,B,\{z_i\}\big)\prod_{i=1}^N\sqrt g \,  \rmd^2z_i
\nonumber\\
&=\log \int e^{-S_\b(\varphi)-S_\f(\psi,\bar\psi)+N\log\int_\Sigma g_{z\bz}^{-1/2}\psi\bar\psi e^{i\varphi/\sqrt\nu}\sqrt g \rmd^2z}\mathcal D_g\varphi\mathcal D\psi D\bar\psi.
\end{align}
We thus split the generating functional into two parts according to the following principle: The first part, labelled as the effective action $S_{\rm eff}$, contains the \emph{anomalous} terms, i.e., nonlocal functionals of $B$ and $R$, which have a local integral form only when expressed in terms of potentials $\lambda$ and $\sigma$. Note that the label ``effective action'' makes sense because these terms are in one-to-one correspondence with the 2+1 Chern-Simons action \cite{KMMW17}, which is the phenomenological effective action that describes the effective topological degrees of freedom of a FQH state. 

To compute $S_{\rm eff}$ explicitly, we note that the normalization constant depends on the metric via the metric-dependence of the determinants as well as the Green's functions in \eq{Vgen}. For the variation of the determinants, we quote the well-known \cite{DMS, DP88} transformation properties of the zeta-determinants of Laplacians \cite[Eq.\ 2.70]{DP88}:
\begin{equation}
\label{trans1}
\frac{\det'\Delta_{g_{\phantom{0}}}}{\det'\Delta_{g_0}} = \exp\left\{-\frac16S_L[g_0,\sigma] \right\}, \qquad
\frac{\det\bp^\dagger\bp|_{g_{\phantom{0}}}}{\det\bp^\dagger\bp|_{g_0}} = \exp\left\{\frac1{12}S_L[g_0,\sigma]\right\},
\end{equation}
where the Liouville action is given by 
\be\label{SL}
S_L[g_0,\sigma]=\frac1{2\pi}\int_\Sigma\left(-\frac14\sigma\Delta_{g_0}\sigma+\frac12R_0\sigma\right)\sqrt{g_0} \, \rmd^2z,
\ee
where $R_0$ is the scalar curvature of $g_0$. For the variations of the Green's functions, we use the expressions from Ref.~\cite[Eq. 4.4--4.6]{FK14}. 
Putting everything together, the effective action becomes 
\be\label{Seff}
S_{\rm eff}[g_0, B_0, g, B] = \nu\NPhi^2 S_2[g_0,B_0,\lambda]-\frac12Q\sqrt\nu\NPhi S_1[g_0,B_0,\sigma,\lambda]+ \left(\frac{1-3Q^2}{12} + \frac{1}{24} \right)
S_L[g_0,\sigma].
\ee
This is in the form of \cite[Eq.\ 5.10]{K16}, with the addition of $\frac{1}{24} S_L[g_0, \sigma]$, which is the contribution to the gravitational anomaly from Majorana fermions. The remaining action functionals read 
\begin{align}\label{S1S2}
\nonumber
S_1[g_0,B_0,\sigma,\lambda] &= \frac1{2\pi}\int_\Sigma\left(-\frac12\lambda R_0
+\frac1{\NPhi} B_0\sigma+\frac12\sigma\Delta_0\lambda\right)\sqrt{g_0} \rmd^2z, \\ 
S_2[g_0,B_0,\lambda] &= \frac1{2\pi}\int_\Sigma\left(\frac14\lambda\Delta_0\lambda
+\frac1{\NPhi} B_0\lambda\right)\sqrt{g_0} \rmd^2z.
\end{align}

\subsection{Observables} 
We next use the effective action to compute the expectation values of various observables. Note that the expectation value of an operator $\mathfrak{O}$ with respect to the Pfaffian state is defined as 
\be\label{Oexpt}
   \exptPf{\mathfrak{O}}\equiv\int_{\Sigma^N} \mathfrak{O} \, \abs{\Psi_{\pf} \zs}^2 \, \prod_{i=1}^N \sqrt{g} \,  \rmd^2z_i,
\ee
and a suitable choice of $\mathfrak{O}$ can be used to compute various transport coefficients. For instance, the charge density, given by 
\be\label{rhocorr}
\exptPf{\rho(\vz)}= \int_{\Sigma^N} \rho(\vz)\, 
 |\Psi_\pf \zs|^2 
\prod_{i=1}^N\sqrt g \,  \rmd^2z_i = \frac1{Z[g,B]}\int_{\Sigma^N} \rho(\vz)\, 
% |\Psi_\pf (\{\vz_i\})|^2 
\mathcal{V}(g,B, \{\vz_i\})
\prod_{i=1}^N\sqrt g \,  \rmd^2z_i,
\ee
is the expectation value of $\rho(\vz)=\sum_{i=1}^N\delta_g(\vz,\vz_i)$. We stress that this corresponds to the density in an inhomogeneous magnetic field on a curved surface with an arbitrary metric $g$. Obviously, $\int\langle\rho(\vz)\rangle\sqrt g \,  \rmd^2z=N$ is the total number of particles. Another quantity of interest is the \emph{static structure factor}, defined on the plane as the Fourier transform of the two point density correlator 
\be 
    S(k) \propto \int_{\mathbb{C}} \exptPf{\rho(\vz) \rho(0)} \e^{-ik(z-w)}  \rmd^2z.
\ee
Since the Fourier transform is not generically defined on arbitrary Riemann surfaces, we shall compute $S(k)$ in the small-$k$ limit using the two-point correlator $\expt{\rho(\vz) \rho(\vw)}$ and a local version of the Fourier transform.

To relate these observables to the normalization constant, we set the magnetic field to a constant value $B_0 = N_\phi$ following the quantization condition of \eq{quant}. This then leads to a relation between $\sigma$ and $\lambda$ as 
\be\label{psisigma}
\sigma(\vz) =\log\left(1+\frac12\Delta_0\lambda(\vz)\right).
\ee 
A variational derivative of \eq{Z0} with respect to $\lambda$ then results in
\be\label{var}
\frac1{\sqrt{g}(\vz)}\frac{\delta}{\delta\lambda(\vz)}\log\frac{Z[g_0,B_0,\sigma,\lambda]}{Z[g_0,B_0]} =
\left(-\NPhi+\frac{1-s}2\Delta_g\right)\exptPf{\rho(\vz)}.
\ee
After another variational derivative, we get the two-point correlator required to compute the static structure factor:
\be\label{ward}
\frac1{\sqrt{g}(\vz)}\frac{\delta}{\delta\lambda(\vz)}\left(\sqrt g(\vw)\exptPf{\rho(\vw)} \!\right)= \left(-\NPhi+\frac{1-s}2\Delta_g^z\right)\exptPf{\rho(\vz)\rho(\vw)}^c,
\ee
where the superscript in $\Delta_g^z$ indicates that it acts on the variable $\vz$ and the connected two point function is defined as
\be\label{2pt}
\exptPf{\rho(\vz)\rho(\vw)}^c = \exptPf{\rho(\vz) \rho(\vw)} - \exptPf{\rho(\vz)}\exptPf{\rho(\vw)}.
\ee
Furthermore, the functional $\mathcal F[g,B]-\mathcal F[g_0,B_0]$ in \eq{ZZ} depends on the metric only through the scalar curvature. By power counting, the first local integral correction is of the form
\be
\mathcal F[g,B]-\mathcal F[g_0,B_0]\sim \frac1N\left(\int_\Sigma R^2\sqrt g \rmd^2z-\int_\Sigma R_0^2\sqrt{g_0} \rmd^2z\right) + \mathrm O(N^{-2}).
\ee
Thus, the $\mathrm O(1)$ and higher contributions to \eq{ZZ} come only from the anomalous terms in $S_{\rm eff}$. This is analogous to the computation in \cite[\S 5]{FK14}, where the saddle point expansion of \eq{F} in $1/N$ reduces to local integrals involving polynomials of $B$ and $R$ and their derivatives e.g., in the integer QHE \cite[Eq.\ 55]{KMMW17} and the Laughlin state \cite[Eq.\ 130]{CLW15}, \cite[Eq. 5.15]{K16}.  

Using \eq{psisigma}, the action functionals in \eq{S1S2} can be identified  as the Mabuchi and the Aubin-Yau functionals given by \cite{FKZ12, FK14} 
\begin{align}
\nonumber 
S_{\rm M}[g_0,\lambda] = S_1[g_0,\NPhi,\sigma(\lambda),\lambda]
&= \frac1{2\pi}\int_\Sigma\left(-\frac12\lambda R_0
+\big(1+\frac12\Delta_0\lambda\big)\log\big(1+\frac12\Delta_0\lambda\big) \right)\sqrt{g_0} \rmd^2z,  \\
S_{\rm AY}[g_0,\lambda] = S_2[g_0,\NPhi,\lambda]
% = S_2(g_0,\NPhi,\lambda)
&= \frac1{2\pi}\int_\Sigma\left(\frac14\lambda\Delta_0\lambda
+\lambda\right)\sqrt{g_0} \rmd^2z.
\end{align}
Thus, the effective action for a constant magnetic field is 
\be\label{Seff2}
    S_{\rm eff}[g_0, \lambda] = \nu\NPhi S_{\rm AY}[g_0,\lambda] - \frac12Q\sqrt\nu\NPhi^2 S_{\rm M}[g_0,\lambda] + \left(\frac{1-3Q^2}{12} + \frac{1}{24} \right) S_{\rm L}[g_0,\sigma(\lambda)].
\ee

\subsection{Explicit computations}
We finally compute the explicit form of the electron density and the static structure factor for the Pfaffian state on the sphere using Eqns~\eqref{var} and \eqref{ward}. 

\subsubsection{One-point function}
The variation of various functionals in \eq{Seff2} are given by \be\label{GravAn}
    \mathfrak{d}_\lambda S_{\rm AY}[g_0, \lambda] = \frac{1}{2\pi}, \qquad 
    \mathfrak{d}_\lambda S_{\rm M}[g_0, \lambda] = -\frac{1}{4\pi} R, \qquad 
    \mathfrak{d}_\lambda S_{\rm L}[g_0, \sigma] = \frac{1}{8\pi} \Delta_{g} R,
\ee
where we have defined $\mathfrak{d}_\lambda = g^{-1/2}(\vz) \,  \delta/\delta \lambda(\vz)$. Thus, using \eq{var}, we derive the large-$N$ expansion for the density as 
\be\label{TYZ}
\expt{\rho(\vz)}_{\rm Pfaffian} = \frac{\nu}{2\pi}\NPhi+\frac{\sqrt\nu \,  Q}{8\pi}R+\frac1{8\pi\NPhi}\left[\frac18+ \frac{\sqrt\nu \, Q}{4} \bigg(1-\frac1\nu\bigg)\right]\Delta_g R+\mathrm O(\NPhi^{-2}),
\ee
where the corrections are higher-order covariant derivatives of the curvature, as follows from implicit diffeomorphism invariance. This is to be compared with the corresponding formula for the Laughlin state with filling fraction $\nu$ \cite{CLW15}:
\be\label{CLWEq}
\expt{\rho(\vz)}_{\rm Laughlin} =\frac{\nu}{2\pi}\NPhi+\frac{\sqrt\nu \,  \widetilde{Q}}{8\pi}R+\frac1{8\pi\NPhi}\left[\frac1{12}+\frac{\sqrt\nu \, \widetilde{Q}}{4}\bigg(2-\frac1\nu\bigg)\right]\Delta_g R+\mathrm O(\NPhi^{-2}),
\ee
where $\widetilde{Q}=1/\sqrt\nu-2s\sqrt\nu=Q-\sqrt\nu$. 

We point out two central differences  between the density of the Laughlin and the Pfaffian states. First, the background charge $Q$ for the Pfaffian state is shifted by $\sqrt{\nu}$ due to the intrinsic $1/2$-spin of fermionic operators. Second, the coefficients $1/8$ and $1/12$ in the $\mathrm{O}(N_\phi^{-1})$ term are due to a difference in central charge between the corresponding CFTs.  For the Laughlin state, we have a single scalar boson with central charge $c=1$, while for the Pfaffian state,  we get $c=3/2$, with the additional $1/2$ coming from the Majorana fermions in the Ising CFT. The central charge appears in the gravitational anomaly contribution $\frac{c}{12} \Delta_g R$, which originates from the zeta-function regularized operator determinants defined in \eq{trans1}. 
% $1/2$ factor in the third term in \eq{TYZ} should be read as $c/3$, where is contribution to the gravitational anomaly coming from the zeta-determinants . For the Laughlin state the same contribution in \eq{CLWEq} corresponds to $c=1$.

\subsubsection{Two-point function}
A formal expression for the two point function defined in \eq{2pt} can be computed by differentiating \eq{TYZ} with respect to $\lambda(\vw)$ to get 
\begin{multline}
\frac1{\sqrt{g}(\vw)}\frac{\delta}{\delta\lambda(\vw)}\big(\sqrt g(\vz)\exptPf{\rho(\vz)}\big)= \frac1{4\pi}\nu\NPhi\,\Delta_g \delta_g(\vz,\vw) -\frac1{16\pi}\sqrt\nu Q\,\Delta_g^2 \delta_g(\vz,\vw) \\
-\frac1{64\pi\NPhi}\left(\frac12+\sqrt\nu Q\bigg(1-\frac1\nu\bigg)\right)\left(\Delta_g\big(R\Delta_g\delta_g(\vz,\vw)\big)+\Delta_g^3\delta_g(\vz,\vw)\right)+\mathrm O(\NPhi^{-2}).
\end{multline}
Integrating \eq{ward}, we arrive at
\begin{align}\label{2pt2}
\exptPf{\rho(\vz)\rho(\vw)}^c &= -\frac\nu{4\pi}\Delta_g\delta_g(\vz,\vw)+\frac1{16\pi\NPhi}(1-\nu)(\Delta_g)^2\delta_g(\vz,\vw) \nonumber \\
&\quad + \frac1{64\pi\NPhi^2}\left(\frac12+\sqrt\nu Q\bigg(1-\frac1\nu\bigg)\right)\Delta_g\big(R\Delta_g\delta_g(\vz,\vw)\big) \nonumber \\
&\quad + \frac1{128\pi\NPhi^2}\left(1-2\nu-\frac2\nu\right)\Delta_g^3\delta_g(\vz,\vw)+\mathrm O (N_\phi^{-3} ).
\end{align}
This is a formal expansion with singular coefficients, so the scope of its validity, if any, is limited. Its main use is to extract the static structure factor at small momenta, defined as follows. Consider the setting where the curvature of the metric $g$ vanishes in a macroscopic part of the surface $\Sigma$, i.e., $R=0$ in a ball $B_L(0)$ centered at the point $z=0$ of radius $L\geq\ell$, where $\ell=\sqrt{\hbar/eB}$ is the magnetic length. In other words, the metric on the ball $B_L(\vz)$ is the flat metric $\rmd s^2=|\rmd z|^2$. Then we position the point $w=x_1+ix_2$ in the annulus $\ell\leq|w-z|<L$ and define the Fourier transform of \eq{2pt} in this Euclidean geometry as
\be\label{stat}
    S(k)=\frac{2\pi}N\int_{\mathbb R^2}e^{-i \sqrt\NPhi (k_1x_1+k_2x_2)}\exptPf{\rho(0)\rho(x)}^c \rmd^2x
\ee
The role of the factor $\sqrt\NPhi$ in the exponent is to effectively blow up this geometry from the ball $B_{L}(0)\to B_{\sqrt\NPhi L}(0)$ to the whole  plane $\mathbb R^2$.
Using \eq{2pt2} and the relation $N=\nu\NPhi$ valid on the plane, we arrive at  
\be\label{Sk}
S_{\rm Pfaffian}(k)=\frac{k^2}2+\frac{1-\nu}{8\nu}k^4+\frac{(1-2\nu)(2-\nu)}{64\nu^2}k^6+ \mathrm O(k^8). 
\ee
For comparison we quote corresponding result for the Laughlin state \cite{CLW15,KMST00}
\be
S_{\rm Laughlin}(k)=\frac{k^2}2+\frac{1-2\nu}{8\nu}k^4+\frac{(1-3\nu)(3-4\nu)}{96\nu^2}k^6+ \mathrm O(k^8)
\ee
As a consistency check, we note that \eq{Sk} is independent of the gravitational spin, as it is be expected since $s$ drops out from the definition in \eq{Psi2} when the metric is Euclidean. Another similarity is that in both cases the leading correction vanishes for the bosonic state, i.e., $\nu=1/2$ for the Laughlin state and $\nu=1$ for the Pfaffian state, while the subleading correction vanishes for the first non-trivial fermionic FQHE state, i.e., $\nu=1/3$ for the Laughlin state and $\nu=1/2$ for the Pfaffian state.

\section{Conclusions}
\label{sec:conc}
In this article, we generalize the Pfaffian state to Riemann surfaces with arbitrary curved metrics and inhomogeneous magnetic fields and compute the generating functional, i.e., the logarithm of the $L^2$--norm of the wave function, in terms of the background metric and the magnetic field. The essential tool for this computation is the description of the Pfaffian state as a correlation function in the Ising CFT \cite{MR91}. This CFT can be readily generalized to curved backgrounds and the correlation functions of a string of $N$ electron operators -- corresponding to having $N$ particles -- can be readily computed in a $1/N$ expansion using the path integral formalism. Setting the magnetic field to a constant, the variations of the generating functional are further used to compute the electron density as well as the static structure factor in the long distance limit. 

The trial wave functions proposed for various fractional quantum Hall plateaux are not expected be the exact ground states of a realistic interacting fermionic Hamiltonian. However, they are supposed to capture the universal features of the state physically realized in the system and thus yield aeffective action which correctly encodes the response to external fields. The comparison of our results for the electron density of the Pfaffian state with that obtained earlier for the Laughlin state \cite{CLW15, FK14} clearly exhibits this feature, where the terms at $\mathrm O(N)$ and $\mathrm O(1)$ depend only on the global properties, \emph{viz}, the filling fraction and the background charge, respectively. Physically, they are related to the Hall conductance and the Hall viscosity, respectively \cite{CLW15}. The $\mathrm O(N^{-1})$ term, on the other hand, captures the gravitational anomaly and hence the central charge of the FQHE state. All three terms correspond to the anomalous part of the full effective action.

In this article, we have focused on a manifold with genus zero, i.e, a 2-sphere. However, it is known that the transport coefficients for the quantum Hall states appear as Chern classes of vector bundles corresponding to the lowest Landau levels on moduli spaces of surfaces with genus $\g>0$ 
\cite{ASZ94,ASZ95,R09,KW15,BR15-2,KMMW17}. In order to understand these Chern classes, it would be interesting to construct the Pfaffian states on higher-genus Riemann surfaces along the lines of the Laughlin case \cite{K17}. We note that much is already known about the Pfaffian states on higher-genus surfaces, including topological degeneracy formulas \cite{RG00,OKSNT07,ABKW08} and $\g=1$ wave functions \cite{GWW92-2,CS07}. The transport coefficients on curved spaces, including the central charge, have also been recently measured experimentally in photonic systems with synthetic Landau levels \cite{SRGSS16,SCCGS19}. Thus, the study of FQH trial wave functions on nontrivial backgrounds still holds many results of both theoretical and experimental interest.  \\ 

 \noindent {\bf Acknowledgements}. 
We would like to thank E. Ardonne, T.~Can, O.~Golan, T.~Hansson, M. Hermanns, C.~Hickey, H.~Legg, N.~Nemkov and P.~Wiegmann 
for useful discussions on this paper and related topics. We also thank A.~Gromov for sharing his preprint \cite{GMR19} and for useful discussions. VD was funded by the \textit{Deutsche Forschungsgemeinschaft} (DFG, German Research Foundation) -- \textit{Projektnummer} 277101999 -- TRR 183 (project B03). 
 SK was partially funded by the \textit{Deutsche Forschungsgemeinschaft} (DFG) -- \textit{Projektnummer} 376817586 --  CRC/TRR 191, RFBR grant 17-01-00585, as well as by the QM2 collaboration and the UoC Forum "Classical and quantum dynamics of interacting particle systems" of the Institutional Strategy of the University of Cologne within the German Excellence Initiative. 

\pagebreak

\bibliographystyle{unsrt_vd}
\bibliography{pfaffian}

\end{document}